\documentclass[english]{article}

\usepackage[preprint, nonatbib]{neurips_2024}
\usepackage[utf8]{inputenc} 
\usepackage[T1]{fontenc}    
\usepackage{csquotes}
\usepackage[
    backend=biber,
    style=numeric,
    citestyle=authoryear,
    maxcitenames=2,
    maxbibnames=5,
]{biblatex}
\newcommand{\citet}{\textcite}
\newcommand{\citep}{\parencite}

\usepackage{hyperref}       
\usepackage{url}            
\usepackage{booktabs}       
\usepackage{amsfonts}       
\usepackage{nicefrac}       
\usepackage{microtype}      
\usepackage{xcolor}         
\usepackage{algorithm2e}

\usepackage{ifthen}
\usepackage{color}
\usepackage{graphics,graphicx,epsfig}
\usepackage{epsf,epstopdf,wrapfig}
\usepackage{amssymb,amsfonts,amsmath,babel,mathtools,bm}
\usepackage[export]{adjustbox}
\usepackage{subfigure}

\usepackage{amsthm}

\usepackage{multirow}

\definecolor{ocre}{RGB}{10,100,185}
\usepackage[normalem]{ulem}

\newcommand{\norm}[1]{\left\lVert#1\right\rVert}

\def\eqref#1{equation~\ref{#1}}

\def\1{\bm{1}}

\def\eps{{\epsilon}}

\DeclareMathAlphabet{\mathsfit}{\encodingdefault}{\sfdefault}{m}{sl}
\SetMathAlphabet{\mathsfit}{bold}{\encodingdefault}{\sfdefault}{bx}{n}

\theoremstyle{plain}

\theoremstyle{definition}

\theoremstyle{remark}

\usepackage{soul}

\usepackage[]{hyperref}
\usepackage[capitalize,noabbrev]{cleveref}

\newcommand{\GM}[1]{{\color{black} #1}}

\newcommand{\EQ}{\begin{equation}}
\newcommand{\EE}{\end{equation}}
\newcommand{\EQA}{\begin{eqnarray}}
\newcommand{\EEA}{\end{eqnarray}}

\title{Loop-Diffusion: an equivariant diffusion model for designing and scoring protein loops}

\author{%
  Kevin Borisiak\\
  Department of Physics, University of Washington \\
  \texttt{kborisia@uw.edu} \\
  \And
  Gian Marco Visani\thanks{Correspondence should be addressed to Gian Marco Visani (gvisan01@cs.washington.edu) and Armita Nourmohammad (armita@uw.edu).}\addtocounter{footnote}{-1}\addtocounter{Hfootnote}{-1} \\
  Paul G. Allen School of Computer Science and Engineering, University of Washington\\
  \texttt{gvisan01@cs.washington.edu} \\
  \And
  Armita Nourmohammad\footnotemark \\
  Department of Physics, University of Washington \\
  Department of Applied Mathematics, University of Washington \\
  Paul G. Allen School of Computer Science and Engineerings, University of Washington \\
  Fred Hutch Cancer Research Center, Seattle, WA \\
  \texttt{armita@uw.edu}
}

\begin{document}

\maketitle

\begin{abstract}
Predicting protein functional characteristics from structure remains a central problem in protein science, with broad implications from understanding the mechanisms of disease to designing novel therapeutics. Unfortunately, current machine learning methods are limited by scarce and biased experimental data, and physics-based methods are either too slow to be useful, or too simplified to be accurate. In this work, we present Loop-Diffusion, an energy based diffusion model which leverages a dataset of general protein loops from the entire protein universe to learn an energy function that generalizes to functional prediction tasks. We evaluate Loop-Diffusion's performance on scoring TCR-pMHC interfaces and demonstrate state-of-the-art results in recognizing binding-enhancing mutations. 

\end{abstract}

\section{Introduction and Related Work}\label{intro}
Predicting the functional characteristics of proteins has been a central goal in protein science. Within structural biology, the sequence-structure-function paradigm is qualitatively well understood, but rigorous quantitative predictions remain out of grasp. Understanding the relationship between a protein's structure and its function is particularly elusive due to the large number of degrees of freedom involved in determining a protein's structure and its complex dynamics over a wide range of time scales.
This is further exacerbated by the scarcity of experimental data for functional measurements, which is expensive to generate, and  is often noisy, biased, and riddled with batch effects. An important challenge is modeling the interaction between T-cell receptors (TCRs) and peptide-MHC (pMHC) antigens, which could enable the design of novel antigen-specific immune receptors for cancer immunotherapy~\citep{cappell_long-term_2023, cameron_identification_2013} and autoimmune disease prevention~\citep{bjornevik_longitudinal_2022}.  However, computational models in this domain are limited by  data quality.
For example, sequence data for paired TCR-pMHC complexes is limited, highly biased, and skewed toward specific subsets~\citep{shugay_vdjdb_2018}. Additionally, structural information is available for only a few hundred experimentally resolved TCR-pMHC complexes, and computational protein-folding models often lack reliability in this domain, making it difficult to create robust and generalizable models.\\
\\
\GM{Traditionally, molecular dynamics simulations~\citep{ayaz_structural_2023} and empirical energy functions~\citep{chaudhury_pyrosetta_2010} have been used to model protein-protein interactions in a data-free way, but the former are too slow to be useful at scale, and the latter often lack accuracy.
Machine learning is an attractive alternative, but currently available data is unsuitable for supervised learning.}
To ameliorate the data-scarcity issue, a growing body of work has attempted to use unsupervised learning and zero-shot protocols to infer biophysical energy functions for proteins. For example,~\citet{roney_state---art_2022} showed that the output confidence score of a structure prediction model such as AlphaFold~\citep{jumper_highly_2021} can be used to distinguish real protein structures from ``decoy" structures, suggesting that the model has learned some physical potential around the equilibrium structure. Unfortunately, this approach has been less successful on more nuanced tasks:~\citep{pak_using_2023} found that AlphaFold output metrics do not correlate with change of protein stability upon mutations. Other work has sought to augment AlphaFold with domain-specific knowledge to improve its structure predictions and scoring ability, for example for TCR-pMHC complexes~\citep{bradley_structure-based_2023}. Recently, generative models have exploded in popularity owing to their success in natural language processing and computer vision. At their core, generative models seek to learn and sample from the probability distribution of training examples. In equilibrium physical systems, this distribution is intricately tied to the energy of a system through the Boltzmann distribution. Indeed, several works have shown that generative models can be used as zero-shot estimators of energy-based functional quantities like protein stability~\citep{pun_learning_2024, visani_hermes_2024, meier_language_2021} and protein-protein binding affinity~\citep{visani_hermes_2024, jin_dsmbind_2023}. Design choices of the input space and careful curation of the training data are crucial factors that determine the types of quantities a generative model can capture and its overall performance. For example,~\citep{meier_language_2021} used the log-likelihood of a model trained to predict masked amino-acid identities given contextual protein sequence to infer mutation effects on protein function, noting that using training sequences at a higher similarity cutoff improved the zero-shot performance of mutation effects. Similarly,~\citep{pun_learning_2024, visani_hermes_2024} trained models to predict masked amino-acid identities given a contextual atomic structure, and used its log-likelihood to infer mutations' effects on protein stability as well as protein-protein binding.~\citep{jin_dsmbind_2023} instead trained an energy-based model with score matching to score protein-protein interfaces, finding that the learned scores correlate well with the experimentally measured affinity between the binding partners.\\
\\
Protein structures are multi-scaled. Contiguous chunks of amino-acids within the protein's sequence form well defined structural motifs that are conserved across proteins. Between these motifs lie loops, regions with particularly high levels of thermal motion. While the more ordered motifs form the topological structure of the protein, the active regions responsible for protein function  often contain disordered loops, such as the CDR3 loops of immune receptors and the peptide antigens within TCR-pMHC complexes, and the CDR3 loops of antibodies. In this work, we model protein loops to capture the biophysical interactions determining the activity and affinity of loops. We hypothesize that irrespective of their activities, data on loops in their structural contexts should inform the biophysical interactions that sustain a loop in the structure, and ultimately determine its function. Therefore, we propose to leverage the large set of general loops in proteins to learn a model that can score and design active loops, such as CDR3s, peptides, and more. To do so, we present Loop-Diffusion, an energy-based diffusion model trained on 433k atomic neighborhoods surrounding loops of various lengths, extracted from 20k non-redundant protein structures. Loop-Diffusion is trained to generate valid  atom configurations for loops within a fixed local environment. 
With its  energy-based architecture, Loop-Diffusion can be easily used to score loop configurations within their environment. We evaluate the ability of Loop-Diffusion to score mutations on peptides and CDR3 loops within TCR-pMHC interfaces, demonstrating that it achieves state-of-the-art results at recognizing binding-enhancing mutations compared to other unsupervised models from the literature.

\section{Methods}

\subsection{Structure Preprocessing and Loop Extraction}\label{loop_extraction}
\begin{figure}
  \centering
  \includegraphics[width=1.0\linewidth]{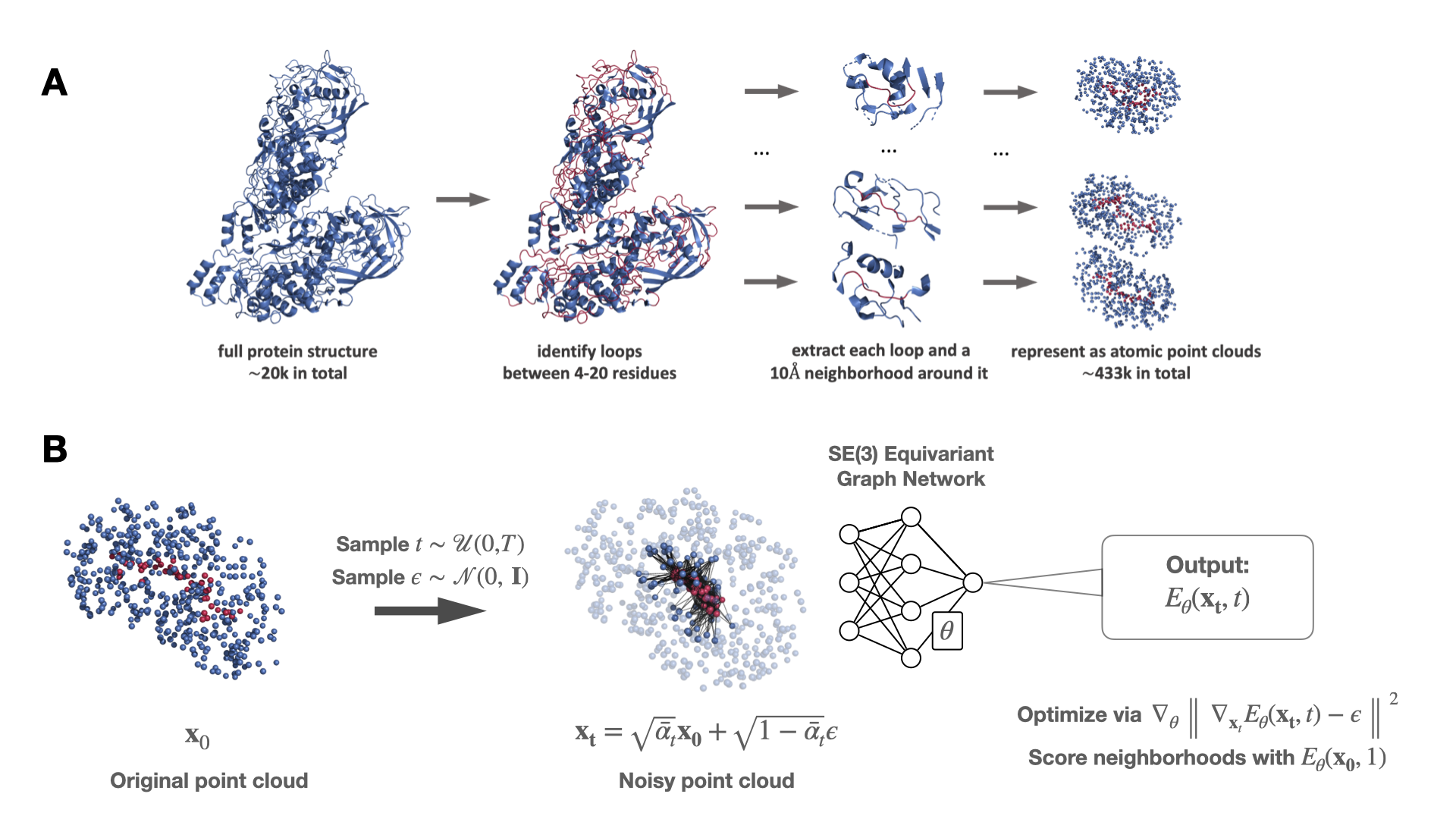}

  \caption{\GM{\textbf{Schematic of Loop-Diffusion.} \textbf{A)} Loop extraction pipeline. We consider all loops of length between 4 and 20 residues, and atoms in their 10 \r{A} neighborhoods. \textbf{B)} Loop-Diffusion is an energy-based model trained with the DDPM objective.}}
  \label{fig:data_pipeline_and_model}
\end{figure}

We use 20k protein structures from the ProteinNet split of CASP12 at 30\% similarity cutoff~\citep{alquraishi_proteinnet_2019}. From a protein structure file, we extract a set of neighborhoods, which are a subset of the full protein structure. We define the neighborhoods as follows. We identify loop residues using the DSSP algorithm within PyRosetta~\citep{chaudhury_pyrosetta_2010}, and identify loops as contiguous sets of loop with length ranging from  4 to 20 amino acids; we show the distribution of loop lengths in our training set in Figure~\ref{fig:loop_length_distribution}. We then define a loop's neighborhood as all atoms within the 10 \r{A} radius of the loop residues' alpha carbons ($\alpha$-C's). Atoms that belong to the loop residues are marked as loop atoms, while the rest of the atoms are marked as the environment (Figure~\ref{fig:data_pipeline_and_model}). In addition to atomic coordinates, we save each atom's element type as well as its partial charge computed by PyRosetta. We omit hydrogens to save compute.

\subsection{Energy Based Diffusion Model}
Our model aims to leverage the information contained within the distribution of loop conformations within protein structures to learn a useful energy function for downstream tasks.
We assume that the loop conformations we observe in crystal structures lie at a local minima of some energy landscape, and that the probability of observing a loop conformation $\mathbf{x}$ is given by a Boltzmann Distribution: $p(\mathbf{x}) = {(e^{- E(\mathbf{x})/kT})}/{Z} \label{Boltzmann}$; where $E(\mathbf{x})$ is some unknown energy function containing physical interactions between the constituents of the loop and its environment, \GM{$Z$ is an unknown normalizing constant, $k$ is the Boltzmann constant, and $T$ is temperature (assumed constant throughout this work). We train a neural network to estimate the energy function using the Denoising Diffusion Probabilistic Model (DDPM) objective, described by~\citet{ho_denoising_2020}.}




\subsubsection{The  Objective of the Denoising Diffusion Probabilistic Model (DDPM)}
The diffusion framework described by \citet{ho_denoising_2020} has three key components: (i) the forward process, (ii) the reverse process, and (iii) the optimization objective. In this work, we focus on the forward process and the optimization objective, as these are the components required to train an energy-based model. The forward process is defined by a fixed Markov chain that gradually adds noise to the data according to a fixed variance schedule $\beta_1, ... ,\beta_T$:
\begin{equation}
    p(\mathbf{x}_{1:T}|\mathbf{x}_{0}) := \prod^{T}_{t=1}p(\mathbf{x}_t|\mathbf{x}_{t-1}), \quad\quad p(\mathbf{x}_t|\mathbf{x}_{t-1}) = \mathcal{N}(\mathbf{x}_t; \sqrt{1-\beta_t}\,\mathbf{x}_{t-1},\, \beta_t \mathbf{I}). \label{ForwardMarkov}
\end{equation}

where $p(x_t | x_{t-1})$ represents the conditional probability distribution of the state at time $t$ given the state at the previous time step, and has a Guassian form with mean $\sqrt{1-\beta_t}\,\mathbf{x}_{t-1}$ and variance $\beta_t  \mathbf{I}$. Using the notation $\alpha_t := 1-\beta_t$ and $\bar{\alpha}_t = \prod^t_{s=1}\alpha_s$, we can sample a noisy ${\bf x}_t$ starting from data ${\bf x}_0 \sim p_{data}$ using $p_{\alpha_{t}}(\mathbf{x}_t|\mathbf{x_0}) = \mathcal{N}(\mathbf{x}_t; \sqrt{\bar{\alpha}_t}\mathbf{x}_0, \, (1-\bar{\alpha}_t)\mathbf{I})$, or equivalently generating the data at time $t$ as $ \mathbf{x}_t = \sqrt{\bar{\alpha}_t}\mathbf{x}_0 + \sqrt{1-\bar{\alpha}_t}\mathbf{\epsilon}$, with  $ \mathbf{\epsilon} \sim \mathcal{N}(0, \mathbf{I})$.


The goal of DDPM is to learn a parametric model of the inverse process $p_\theta(\mathbf{{x}_{t-1}|\mathbf{x}_t})$ so that, starting from pure noise $\mathbf{x}_T$, one can generate samples from the true data distribution by running the inverse of \ref{ForwardMarkov}. Such a model can be trained by optimizing the variational bound on the negative log likelihood. For the author's \citet{ho_denoising_2020} choice of parameterization of model, this yields the following objective function:


\begin{equation}
    \mathcal{L}_{DDPM} = \mathcal{L}_0 + \sum_{t=1}^{T-1}  \mathbb{E}_{\mathbf{x}_0,\mathbf{\epsilon}} \left[ \frac{\beta_t^2}{2\sigma_t^2 \alpha_t
    }\norm{\mathbf{\mathbf{\eps}_\theta}(\mathbf{x}_t,t) - \nabla_{\mathbf{x}_t}\log p_{\alpha_t}(\mathbf{x}_t|\mathbf{x}_0)}^2 \right] + \mathcal{L}_T\label{loss_ddpm}
\end{equation}


The optimal network $\mathbf{\epsilon}_\theta(\mathbf{x}_t,t)$ will approximate the score of the true data distribution, perturbed by some Gaussian kernel $p_{\alpha_t}(\mathbf{x}_t) := \int{p_{data}(\mathbf{x})} p_{\alpha_t}(\mathbf{x}_t|\mathbf{x})d\mathbf{x}$~\citep{song_generative_2020}\citep{ho_denoising_2020}\citep{vincent_connection_2011}. In practice, we parameterize $\mathbf{\epsilon}_\theta(\mathbf{x}_t,t)$ as the negative gradient of our energy model:

\begin{equation}\label{noise_parameterization}
    \mathbf{\epsilon}_\theta(\mathbf{x}_t,t) = -\nabla_{\mathbf{x}_t} E_{\theta}(\mathbf{x}_t,t)
\end{equation}

and adopt $\mathcal{L}_{simple}$ from \citet{ho_denoising_2020}, which drops the scaling coefficient on \ref{loss_ddpm} and replaces the sum over $t$ with a sample from $\mathcal{U}(0,T)$ \ref{Implementation details}. Noting that $\nabla_{\mathbf{x}_t}\log p_{\alpha_t}(\mathbf{x}_t|\mathbf{x}_0) = - \epsilon / \sqrt{1 - \bar{\alpha}_t}$, and $1 / \sqrt{1 - \bar{\alpha}_t}$ is dropped when adopting $\mathcal{L}_{simple}$, our training objective becomes:
\begin{equation}
    \mathcal{L}_\textnormal{Loop-Diffusion} = \mathbb{E}_{t,\mathbf{x}_0,\mathbf{\epsilon}} \left[ \norm{\nabla_{\mathbf{x}_t} E_\theta(\mathbf{x}_t,t) - \mathbf{\epsilon}}^2\right]
    \label{eq:loop_diffusion_loss_function}
\end{equation}

We assume that early in the diffusion process, i.e. small $t$, the distribution $p_{\alpha_t}(\mathbf{x})$ approximates the Boltzmann distribution underlying our data. Thus, we expect that when evaluated on real neighborhoods at the smallest time step $t = 1$, our learned energy model will approximate the true energy $E(\mathbf{x})$ up to a scaling constant:

\begin{equation}
\begin{split}
    -\nabla_{\mathbf{x}} E_{\theta}(\mathbf{x}, 1) &\simeq \nabla_{\mathbf{x}} \log p_{\alpha_1}(\mathbf{x}) \\
    &=  \nabla_{\mathbf{x}} \log e^{ \frac{-E(\mathbf{x})}{kT} } = -kT \nabla_{\mathbf{x}} E(\mathbf{x})
\end{split}
\end{equation}


We implement $E_\theta$ using an equivariant graph convolutional network architecture built with the \texttt{e3nn} library~\citep{geiger_e3nn_2022}. For more details on model implementation and training, see \ref{Implementation details}. 

It is worth noting that \citet{jin_dsmbind_2023} 
 follows a similar approach, employing a simpler Denoising Score Matching (DSM) objective \citep{vincent_connection_2011} to learn $E(\mathbf{x})$. Rather than conditioning on $t$ and learning the full Markov chain, they predict the noise added in a single step without conditioning on the noise level. In theory, this should be equally capable of learning the target Boltzmann distribution, however, DSM models are less effective at sampling from the learned distribution. We believe sampling may be useful, as it would allow us to relax protein structures using our learned energy function, which motivated us to pursue the DDPM approach.

\section{Results}

\begin{figure}[t]
    \centering
    \includegraphics[width=1.0\linewidth]{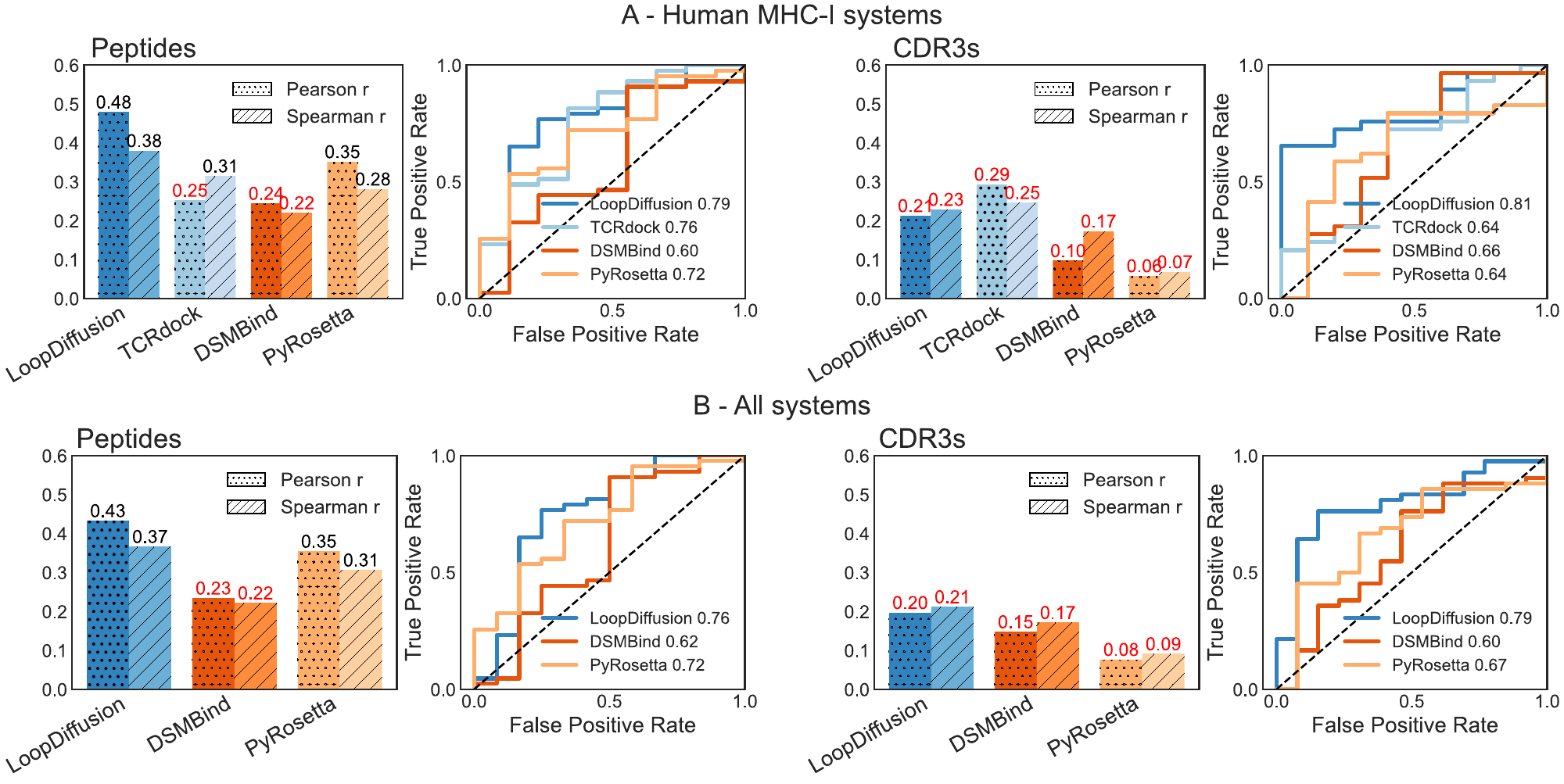}
    \caption{\textbf{Peptides and CDR3 mutational effects on the binding affinity of TCR-pMHC complexes.} The predictions for the effect of mutations on binding affinity in peptides (left) and in the CDR3 loop of TRCs (right) is compared across models  for TCR-pMHC complexes of {\bf(A)} Human MHC-I system (n=52 for peptides, n=37 for CDR3), and {\bf (B)} all the TCR-pMHC complexes in the ATLAS database with a mutation in the peptide or the CDR3 Loop (n=55 for both peptides and CDR3's).
    All panels show correlation coefficients between the experimental data and model predictions (left), and insignificant correlations (p-value>0.05) are indicated in red. All panels show ROC curves and the corresponding AUROC for all models to classify between favorable ($\Delta\Delta G_{binding}\geq0$) and unfavorable ($\Delta\Delta G_{binding}<0$) mutations in each set (right).}
    \label{fig:bar_roc_atlas_mt_struc}
\end{figure}

We evaluated the performance of Loop-Diffusion in predicting the effect of mutations on the binding affinity $\Delta\Delta G$ of the TCR-pMHC complexes. For this task, we leverage the ATLAS dataset~\citep{borrman_atlas_2017}. Due to the nature of our pipeline, we focus specifically on mutations that occur within the loops, specifically on the peptide antigen or the two CDR3 loops of a TCR in the complex. CDR3 loop locations were identified via alignment against annotated TCR sequences, using code from the TCRdock github repository~\citep{bradley_structure-based_2023}. To score a mutation on a loop (either CRD3 or peptide), we extract both the wild-type and the mutant loop neighborhoods from the respective structures; conveniently, ATLAS provides {\em in-silico} generated mutant structures. We evaluate the energy of each structure as the sum of the model node energies evaluated at $t=1$ within the diffusion framework, since at $t=1$ we expect to capture the statistics of the true Boltzmann distribution as closely as possible. We define the model's predicted mutational effect on the binding affinity $\Delta \Delta G$ to be the difference between the predicted energies of the mutant and the wild-type.\\
\\
We compared our approach to three other unsupervised methods, which similar to ours, were not expressively trained to predict mutational effects on binding $\Delta \Delta G$. These methods are: (i) {PyRosetta binding $\Delta\Delta G$~\citep{park_simultaneous_2016}}, computed using our implementation of the ``cartesian-ddG" protocol,  (ii) {TCRdock~\citep{bradley_structure-based_2023}}, which is a protocol enhancing AlphaFold-Multimer's~\citep{evans_protein_2022} ability to predict the structure of TCR-pMHC complexes; it can be used to score TCR-pMHC binding via its predicted alignment error of the interface, and (iii) {DSMBind~\citep{jin_dsmbind_2023}}, which is an energy-based model trained with score matching on protein-protein interfaces. See Section~\ref{sec:baselines_details} for further details on these models.\\
\\
For TCR-pMHC proteins associated with Human MHC Class I, Loop-Diffusion achieves  best correlations to the experimental $\Delta\Delta G$ values for mutations on peptide antigens, and second-best for mutations on CDR3s. In both regions,  Loop-Diffusion shows the best classification accuracy between favorable ($\Delta\Delta G \geq 0$) and unfavorable ($\Delta\Delta G < 0$) mutations, measured by the Area under the Receiver Operating Curve (AUROC); see  Fig.~\ref{fig:bar_roc_atlas_mt_struc}A, and Fig.~\ref{fig:scatterplot_human_mhci_atlas_mt_struc} for the corresponding scatter plots. Moreover, Loop-Diffusion shows best correlations with the experimental data and best classification accuracies for other MHC systems (human MHC II and mouse systems), compared to all other models, except for TCRdock, which cannot be evaluated on this data; see Figs.~\ref{fig:bar_roc_atlas_mt_struc}B,~\ref{fig:scatterplot_all_atlas_mt_struc}.

\section{Discussion}
In this work we have enhanced the power of generative models for zero-shot prediction of protein-protein binding energy by carefully selecting the training data to reflect the distribution of targets of interest. We presented Loop-Diffusion, an energy-based model trained as a DDPM to de-noise loops from the protein universe, and applied it to score mutations within loops at protein's functional interfaces. We tested Loop-Diffusion specifically on CDR3 loops and peptides within TCR-pMHC systems, finding that it is stronger than comparable unsupervised models at identifying binding-enhancing mutations. In future work, we plan on using Loop-Diffusion to score loops within other functional contexts, such as the CDR3 loops of antibodies. Furthermore, we plan on training Loop-Diffusion up to higher noising time-points, thereby granting it generative ability, which can be used for example to generate mutant structures prior to scoring them. On a similar note, we plan on exploring the use of correlated noise structures, so that we can more easily generate valid loop conformations~\citep{jing_eigenfold_2023, jin_dsmbind_2023}.

\section{Acknowledgments}
This work has been supported by the National Institutes of Health MIRA award (R35 GM142795), the CAREER award from the National Science Foundation (grant No: 2045054), A\&S PhD Fellowship Support from the UW Provost, and the Allen School Computer Science \& Engineering Research Fellowship from the Paul G. Allen School of Computer Science \& Engineering at the University of Washington. This work is also supported, in part, through the Departments of Physics and Computer Science and Engineering, and the College of Arts and Sciences at the University of Washington.

\printbibliography

\clearpage{}
\newpage{}

\appendix

\counterwithin{table}{section}
\setcounter{table}{0}
\renewcommand{\thetable}{S\arabic{table}}

\counterwithin{figure}{section}
\setcounter{figure}{0}
\renewcommand{\thefigure}{S\arabic{figure}}

\counterwithin{equation}{section}
\setcounter{equation}{0}
\renewcommand{\theequation}{S\arabic{equation}}


\section{Appendix}


\subsection{Implementation Details}\label{Implementation details}

As input, our model takes a PyTorch Geometric graph of the loop neighborhood, with each atom receiving being assigned to a node with a position value $\mathbf{x}_{pos} \in \mathbb{R}^3$, a feature vector $\mathbf{x}_{feat} \in \mathbb{R}^6$ containing one-hot encoded vector of the 5 atom types it may encounter (N, S, H, C, O) and a single scalar value for the charge, and a one-dimensional attribute vector $\mathbf{z}$ containing a binary value indicating whether the atom belongs to the loop or the environment. During training, we use auto-differentiation to take the gradient of the network energy with respect to the loop node coordinates. 

During training, we append to the node features a 10-dimensional sinusoidal time-embedding, expanding the feature dimension to 16.

For the network, we use the basic graph convolutional network implementation provided within the E3NN package. We experimented with other architectures on a simple n-body force prediction task and found that graph networks outperformed the transformers~\citep{vaswani_attention_2017} and Clebsch-Gordan nets~\citep{kondor_clebsch-gordan_2018} we tried. Additionally, we believe the geometric pairwise interactions encoded by a graph network are more reflective of the underlying physics of the inter-atomic interactions in this problem. E3NN assigns node and edge features according to the Irreducible Representation (irrep) they transform by under SO(3)~\citep{geiger_e3nn_2022}. The irrep is identified by the degree value $l$. $l=0$ corresponds to scalar values, $l=1$ corresponds to vector values, and so on. Within the network, the node and edge features increase up to $l=4$, with a multiplicity of $8$ for each type i.e. each node receives 8 scalars, 8 vectors, 8 traceless symmetric tensors, etc. We found that these values performed well on our baselines during model selection. We use a network depth of 3, 3 radial basis functions for the edge embedding, and 100 radial neurons. 

For our implementation of the diffusion protocol, we choose a linear $\beta_t$ schedule interpolating between $\beta_0 = 0.0001$ and $\beta_{T}= 0.002 $ with $T=2000$. Additionally during training we only sample times from $[0, \frac{T}{2}]$ so the model could better focus on learning the early distribution. We note that omitting the scaling coefficient in \ref{loss_ddpm} is intended to improve sampling quality, which makes it easier to monitor the models learning, however it down-weights the loss at early time steps, which is when the model should be learning the Boltzmann distribution. In future experiments, we would like to train with the weighting coefficient to see if it improves performance on scoring.

\RestyleAlgo{ruled}

\begin{minipage}[t]{0.46\textwidth}
  \begin{algorithm}[H]
    \caption{Training}\label{alg:training}
        \KwData{Loop Neighborhood $\mathbf{x}_0 := [\mathbf{x}_{0, loop}, \mathbf{x}_{0, env}]$}
        \Repeat{converged}
            {sample $\mathbf{x}_0$ from data\\
            sample $t\in [0,\frac{T}{2}]$\\
            sample $\epsilon \in \mathcal{N}(0,\,\mathbf{I})$\\

            add noise to loop coordinates only;\\
            \Indp
                $\mathbf{x}_{t, loop} = \sqrt{\bar{\alpha_t}} \mathbf{x}_{0,loop} + \sqrt{1 - \bar{\alpha_t}} \mathbf{\epsilon}$\\
                $\mathbf{x}_{t, env} = \mathbf{x}_{0, env}$\\
                $\mathbf{x}_t := [\mathbf{x}_{t, loop}, \mathbf{x}_{t, env}]$\\
            \Indm
            take gradient descent step on:\\
            \Indp
            $\nabla_\theta \norm{\nabla_{\mathbf{x}_t} E_\theta(\mathbf{x}_t,t) - \mathbf{\epsilon}}^2$}
  \end{algorithm}
\end{minipage}
   \hfill
\begin{minipage}[t]{0.46\textwidth}
    \begin{algorithm}[H]
        \caption{Inference (specifically mutation scoring)}\label{alg:two}
        \KwData{\\
        Wild Type Neighborhood $\mathbf{x}_{wt}$\\
        Mutant Neighborhood $\mathbf{x}_{mt}$
        }
        \KwResult{\\
        Compute model energies at $t=0$:\\
        \Indp   
            $E_{wt} = E_\theta(\mathbf{x}_{wt},0)$\\
            $E_{mt} = E_\theta(\mathbf{x}_{mt},0)$\\
        \Indm
        $\Delta \Delta G_{pred} = E_{mt} - E_{wt}$
        }
    \end{algorithm}
        
\end{minipage}

\subsection{Baselines details}
\label{sec:baselines_details}

\textbf{PyRosetta~\citep{park_simultaneous_2016}.} We use our implementation of the ``cartesian-ddG" protocol. Specifically, we compute the binding $\Delta G$ of a TCR-pMHC complex as $\Delta G = E_{\text{TCR-pMHC}} - (E_{\text{TCR}} - E_{\text{pMHC}})$, where each energy term $E$ is computed using pyrosetta's cartesian scoring function. We then compute binding $\Delta \Delta G$ simply as $\Delta G_{\text{mt}} - \Delta G_{\text{wt}}$.\\
\\
\textbf{TCRdock~\citep{bradley_structure-based_2023}.} This is an AlphaFold-based algorithm that uses carefully-selected structural templates, alongside considerations about TCR-pMHC's common docking geometries, to enhance AlphaFold-Multimer's~\citep{evans_protein_2022} capabilities on TCR-pMHC structure prediction. TCRdock's PAE score of the TCR-pMHC interface has been shown to have some discriminatory power of correct TCR-pMHC pairings. We thus treat the TCR-pMHC PAE as a binding score, and the difference between mutant and wildtype scores as a predictor of binding $\Delta\Delta G$. As we encountered issues when using TCRdock on complexes with Class II MHCs, we only use TCRdock for predictions with Class I MHCs, and leave the analysis to of MHC-Class II complexes to future work. Notably, as TCRdock is effectively a protein-folding algorithm, it does not rely on the availability of accurate structures, though its performance does deteriorate for TCR-pMHC systems that have low similarity matches among those that have structures in TCRdock's database~\citep{bradley_structure-based_2023}. As all of the TCR-pMHC systems in ATLAS have a wildtype structure in TCRdock's database, the TCRdock scores we show are as good as they can get.\\
\\
\textbf{DSMBind~\citep{jin_dsmbind_2023}.} Similar to Loop-Diffusion, DSMBind is an energy-based model trained with score matching; we use the version of the model trained to score protein-protein interfaces. DSMBind adds noise by randomly roto-translating one of the two binding partners about its center of mass, as well as randomly rotating all the side-chains' orientations. DSMBind also uses ESM2 embeddings~\citep{lin_evolutionary-scale_2023} as features to enhance their predictions, which Loop-Diffusion currently does not.

\begin{figure}
    \centering
    \includegraphics[width=0.8\linewidth]{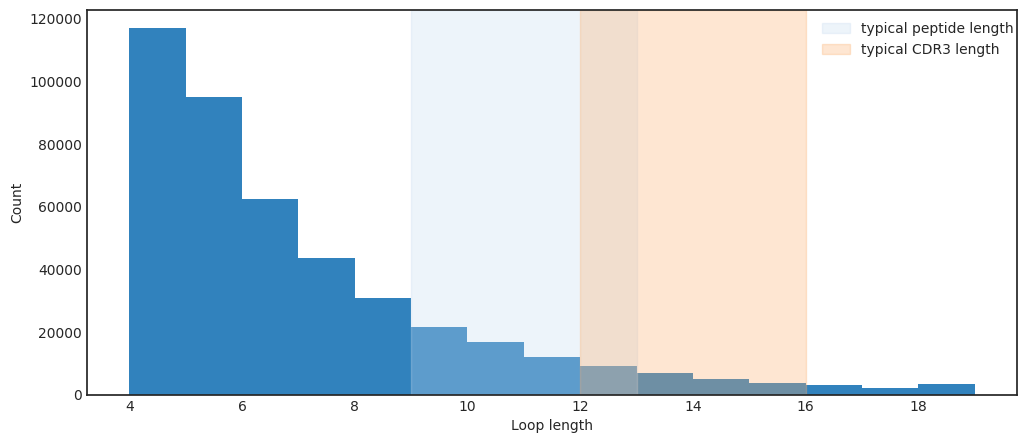}
    \caption{\textbf{Distribution of loop lengths within our dataset extracted from CASP12.}  The peptides in our test dataset range from 9-13 residues in length, which is relatively well represented in our training dataset. CDR3's loops, however, are typically in the range of 12-16 residues in length, which is a more data scarce regime. Future work may attempt to crop the CDR3 loop around the mutation to see if performance is improved.}
    \label{fig:loop_length_distribution}
\end{figure}

\begin{figure}
    \centering
    \includegraphics[width=0.8\linewidth]{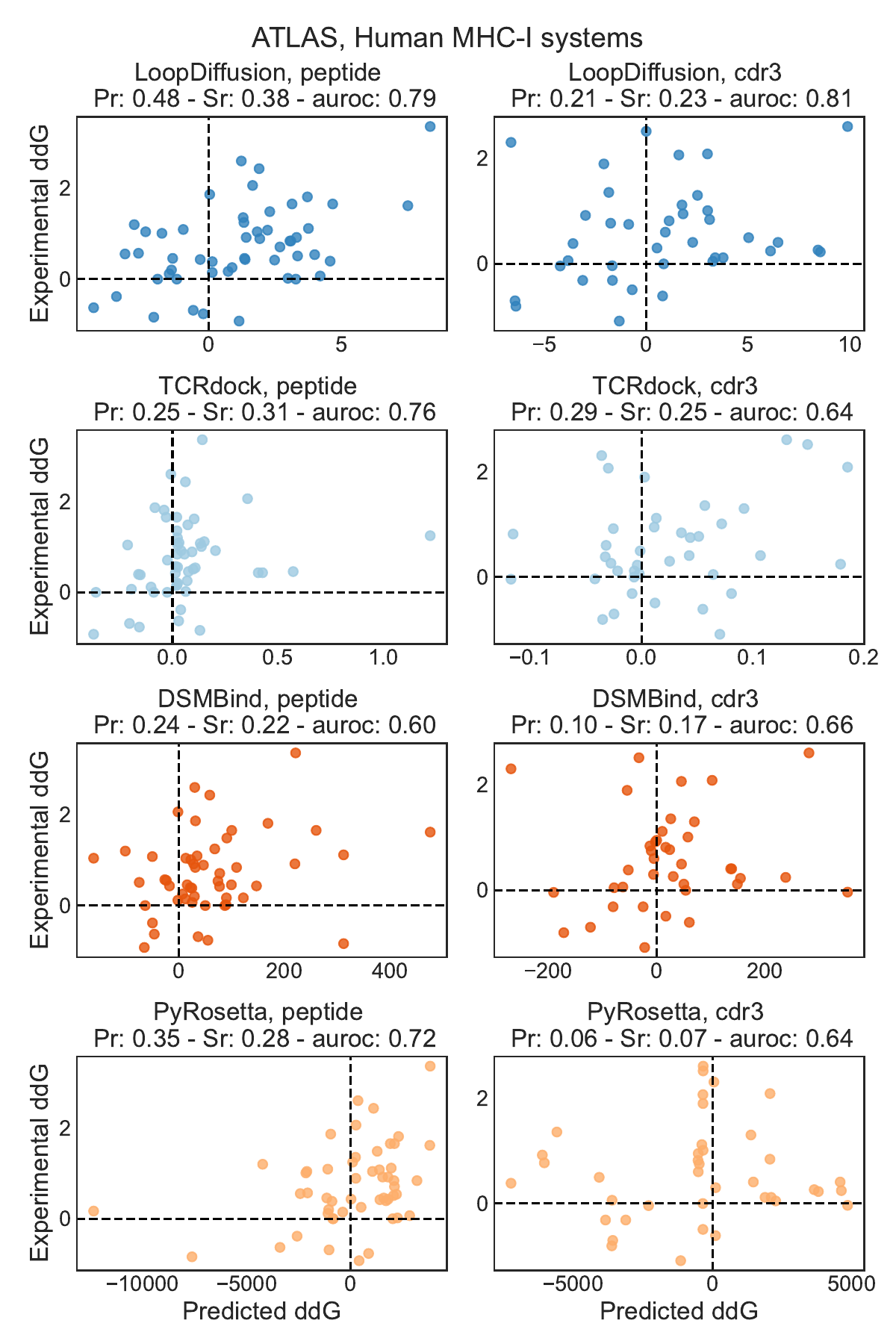}
    \caption{\textbf{Scatterplots of predicted vs. experimental binding $\Delta \Delta G$ on mutations occurring on peptides only (left columns) or one CDR3 only (right column), from the subset of the ATLAS dataset containing only Human MHC Class-I systems.}}
    \label{fig:scatterplot_human_mhci_atlas_mt_struc}
\end{figure}

\begin{figure}
    \centering
    \includegraphics[width=0.8\linewidth]{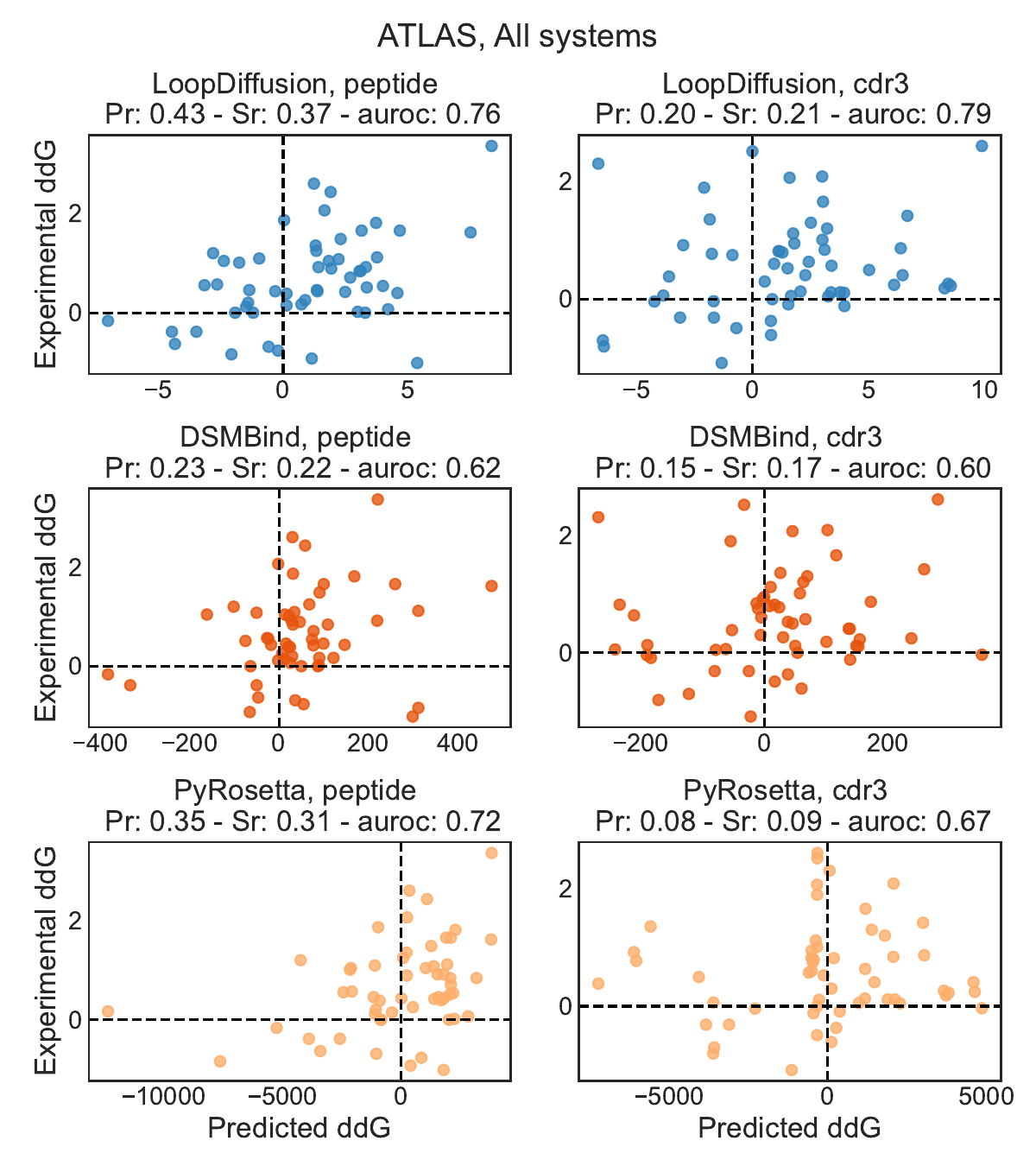}
    \caption{\textbf{Scatterplots of predicted vs. experimental binding $\Delta \Delta G$ on mutations occurring on peptides only (left columns) or one CDR3 only (right column), from the ATLAS dataset.}}
    \label{fig:scatterplot_all_atlas_mt_struc}
\end{figure}

\end{document}